\documentclass[runningheads]{llncs}
\usepackage{subcaption}
\usepackage{booktabs}
\usepackage{multirow} 
\usepackage[T1]{fontenc}
\usepackage[numbers]{natbib}
\usepackage{graphicx}

\begin{document}
\title{Human-in-the-Loop Annotation for Image-Based Engagement Estimation: Assessing the Impact of Model Reliability on Annotation Accuracy}
\titlerunning{Human-in-the-Loop Annotation for Image-Based Engagement Estimation}

\author{
Sahana Yadnakudige Subramanya\inst{1}\orcidID{0009-0002-9504-7479} \and \\
Ko Watanabe\inst{2}\orcidID{0000-0003-0252-1785} \and \\
Andreas Dengel,\inst{2}\orcidID{0000-0002-6100-8255} \and \\
Shoya Ishimaru\inst{3}\orcidID{0000-0002-5374-1510}
}

\authorrunning{Subramanya and Watanabe et al.}

\institute{
RPTU Kaiserslautern-Landau, Kaiserslautern, Germany \and
German Research Center for Artifcial Intelligence(DFKI), Kaiserslautern, Germany \and Osaka Metropolitan University, Osaka, Japan
}
%
\maketitle              
\begin{abstract}
Human-in-the-loop (HITL) frameworks are increasingly recognized for their potential to improve annotation accuracy in emotion estimation systems by combining machine predictions with human expertise. This study focuses on integrating a high-performing image-based emotion model into a HITL annotation framework to evaluate human-machine interaction's collaborative potential and uncover the psychological and practical factors critical to successful collaboration. Specifically, we investigate how varying model reliability and cognitive framing influence human trust, cognitive load, and annotation behavior in HITL systems. We show that model reliability and psychological framing significantly impact annotators’ trust, engagement, and consistency, offering insights into optimizing HITL frameworks. Through three experimental scenarios with 29 participants—baseline model reliability (S1), fabricated errors (S2), and cognitive bias introduced by negative framing—we analyzed behavioral and qualitative data (S3). Reliable predictions (S1) yielded high trust and annotation consistency, while unreliable outputs (S2) induced critical evaluations but increased frustration and response variability. Negative framing (S3) revealed how cognitive bias influenced participants to rate the model as relatable and accurate despite misinformation about its reliability. These findings highlight the importance of reliable machine outputs and psychological factors in shaping effective human-machine collaboration. By leveraging the strengths of both human oversight and automated systems, this study establishes a scalable HITL framework for emotion annotation and sets the stage for broader applications in adaptive learning and human-computer interaction.
\end{abstract}


\section{Introduction}
Emotion recognition, a key area of artificial intelligence (AI), aims to interpret human emotions using data such as facial expressions~\cite{watanabe2021discaas, watanabe2023engauge, chen2022quantitative}, body language~\cite{luo2024emotion}, vocal tones~\cite{el2011survey}, eye-tracking~\cite{john2019pupil, matsuda2018emotour, bhatt2024estimating}, and physiological signals~\cite{tanaka2024concentration, shu2018review}. Advances in machine learning and deep learning have significantly improved the accuracy of emotion recognition systems, enabling applications in diverse fields such as healthcare, entertainment, and human-computer interaction (HCI)~\cite{younis2024machine, erat2024emotion, wu2016review}. Among the various emotional metrics, engagement is a critical factor for interactive applications like virtual assistants, educational platforms, and adaptive systems, where understanding user involvement can enhance responsiveness and effectiveness~\cite{d2013autotutor, bosch2015automatic}. 

Engagement, defined as the level of attention and involvement in a task, offers deeper insights into user behavior. For instance, in e-learning environments, accurate engagement estimation enables dynamic content adjustment, fostering personalized and effective educational experiences~\cite{henrie2015measuring, sharma2015displaying, watanabe2024metacognition}. Despite advances, engagement estimation remains challenging due to individual variability, cultural differences, and the subtle nature of engagement cues such as micro-expressions or fleeting behaviors~\cite{elfenbein2002universality, wang2018micro}. Overfitting on narrow datasets also limits model generalizability across diverse contexts~\cite{yan2019cross}. These challenges highlight the need for innovative approaches that combine automated systems’ scalability with human contextual expertise.

Human-in-the-Loop (HITL) frameworks~\cite{wu2022survey} address these limitations by integrating human expertise into AI-driven processes. HITL systems enable human annotators to refine and correct model predictions, improving annotation accuracy and contextual relevance. This collaborative approach is particularly valuable for emotion recognition. However, little is known about how model reliability affects human behavior in HITL systems, such as how annotators respond to unreliable predictions or how their trust impacts annotation quality. Understanding these dynamics is critical to optimizing human-machine collaboration.

To evaluate the performance of different HITL conditions, we investigated three experimental scenarios (S1-S3): baseline reliability with unaltered predictions (S1), fabricated errors to test responses to unreliable outputs (S2), and cognitive bias induced by framing the model as unreliable despite identical predictions to S1 (S3). This research bridges the gap between automated and manual annotation methods, paving the way for improved engagement estimation in applications such as adaptive learning, virtual environments, and healthcare.
In summary, the contributions of this study are as follows: 

\begin{enumerate}
    \item[C1] We propose an image-based engagement estimation model using in-the-wild facial image dataset, achieving an F1 score of 86\%.
    \item[C2] We develop a HITL application for real-time engagement relabeling.
    \item[C3] We assess human annotation behavior under varying levels of model reliability and analyze the impact of cognitive bias in HITL systems.
\end{enumerate}

\section{Related Work}
\subsection{Emotion Recognition and Engagement Estimation}
\subsubsection{Technical Foundations of Emotion Recognition}
Emotion recognition relies on artificial intelligence (AI) systems to analyze multimodal data and classify emotional states. These systems process raw inputs—such as images, audio, or physiological signals—by extracting features and mapping them to discrete categories (e.g., happiness, anger) or continuous dimensions (e.g., valence, arousal). Early methods employed handcrafted features, such as Gabor filters for facial expressions~\cite{barbu2010gabor} and MFCCs for audio signals~\cite{ali2020mel}, but struggled with the complexity of real-world data. 
Modern systems leverage machine learning and deep learning to learn hierarchical representations automatically. Convolutional neural networks (CNNs) excel at spatial patterns in image data~\cite{debnath2022four}, while recurrent neural networks (RNNs) and transformers effectively handle sequential data, capturing subtle cues like tonal variations or micro-expressions~\cite{lim2016speech, wang2023deep}. These advancements have significantly enhanced the ability of AI systems to detect and interpret emotional signals.

\subsubsection{Engagement as a Focus in Emotion Recognition}
Engagement is a dynamic and context-sensitive emotional state that reflects a user’s attention or involvement. It provides actionable insights for adaptive systems, such as e-learning platforms and human-computer interaction (HCI) frameworks~\cite{d2013autotutor, bosch2015automatic}. 
Estimating engagement involves challenges such as the ambiguity of cues (e.g., gaze, posture, or micro-expressions) and their dependence on context~\cite{markus2014culture}. Multimodal approaches integrating modalities like facial expressions, voice, and physiological data improve accuracy but complicate data synchronization, especially with noisy or incomplete inputs~\cite{zeng2007survey, vairamani2024advancements}.
Engagement also evolves, requiring temporal modeling techniques such as RNNs or transformers to capture patterns and intervene effectively. For instance, in e-learning, systems can detect waning focus and re-engage students through interactive quizzes or tailored content~\cite{karaoglan2022learning}. However, the subtlety of engagement cues, such as slight shifts in head position or blink rates, remains a significant challenge, particularly in low-resolution or noisy data~\cite{oertel2021towards}.

\subsection{Human-in-the-Loop (HITL) Systems}
\subsubsection{Concept and Role of HITL Systems}
HITL systems combine the efficiency of automated models with the contextual understanding of human expertise. Machines process large-scale data but struggle with edge cases or ambiguity, while humans provide intuition and domain knowledge~\cite{amershi2014power}. HITL frameworks enable human intervention during annotation, validation, or refinement stages. For instance, human annotators can correct model misclassifications in emotion recognition by incorporating broader context~\cite{holstein2019improving}. This iterative process improves immediate accuracy and enhances model performance through feedback in subsequent training cycles.

\subsubsection{HITL in Emotion Recognition}
The complexity of emotion recognition, particularly for nuanced states like engagement, makes it an ideal application for HITL frameworks. Automated models often overemphasize specific features, such as gaze, which may be misleading in specific contexts. HITL systems mitigate these issues by enabling human annotators to refine predictions, ensuring accurate and contextually appropriate annotations. This approach is especially valuable for noisy or biased data, such as cross-cultural emotion recognition~\cite{gong2023cross}. HITL frameworks also enhance generalization by incorporating diverse human corrections, enabling models to adapt to various expressions and contexts. This hybrid methodology ensures robustness in real-world applications.

\subsubsection{Advantages of HITL for Engagement Estimation}
HITL frameworks offer significant advantages for engagement estimation by leveraging human expertise to address challenges that automated models often face. Humans can interpret subtle and ambiguous cues, especially in context-dependent scenarios where behavioral signals vary based on individual, cultural, or environmental factors~\cite{salam2022automatic}. By validating and refining model predictions, HITL systems improve annotation accuracy and reliability, ensuring higher-quality data for training and evaluation~\cite{zhang2023labelvizier}. Additionally, human intervention is critical in identifying and mitigating biases in model outputs, resulting in more balanced and equitable predictions~\cite{lukac2023study}. 

HITL systems are particularly beneficial for real-time applications such as adaptive learning and virtual reality (VR) environments. For instance, IoT-based learning platforms leverage human feedback to adjust content dynamically, sustaining user engagement and improving learning outcomes~\cite{taherisadr2023erudite}. Likewise, immersive VR interfaces can adapt to user interactions in real-time, enhancing engagement and overall satisfaction~\cite{yigitbas2021enhancing}.

\section{Methodology}
\subsection{Dataset Description}
This study utilizes the DAiSEE dataset~\cite{gupta2016daisee}, a publicly available resource for emotion recognition in e-learning contexts, capturing natural affective states — engagement, boredom, confusion, and frustration—in real-world settings. This makes it valuable for developing models tailored to educational technology applications.

The dataset includes 9,068 ten-second video snippets from 112 participants, mostly university students aged 18–30, recorded in natural environments like dormitories and libraries. Its varied settings and lighting conditions enhance ecological validity, supporting models that generalize across user environments. Each video is annotated with four affective states on a four-point scale, capturing subtle emotional variations critical for adaptive systems.

The videos are high-resolution webcam recordings (1920x1080 at 30 fps) that capture facial expressions, body language, and non-verbal cues essential for emotion recognition. Participants viewed educational and recreational content, eliciting authentic emotional responses representative of real-world e-learning scenarios.

Annotations were crowdsourced and validated using a gold-standard subset labeled by expert psychologists. The Dawid-Skene algorithm consolidated annotations by weighting annotator reliability, ensuring consistent, high-quality labels for model training and evaluation.

The dataset is split into training, validation, and test sets in a 60:20:20 ratio, with balanced distributions of gender and affective states. This ensures consistent evaluation and supports the development of models that generalize effectively across diverse populations and learning environments.

In summary, DAiSEE provides a rich, well-annotated resource for multi-level engagement recognition in realistic e-learning settings. It combines detailed annotations, environmental diversity, and demographic balance to support advancements in engagement estimation models.

\subsection{Video-Based Emotion Detection}
The MobileNetV2 architecture was chosen as the base model due to its balance between computational efficiency and high performance. It is particularly suitable for video-based data where the sequential nature of frames imposes significant computational demands. Its lightweight structure enables real-time processing, critical for emotion detection in dynamic environments.

\begin{figure}[t]
    \centering
    \includegraphics[width=0.7\textwidth]{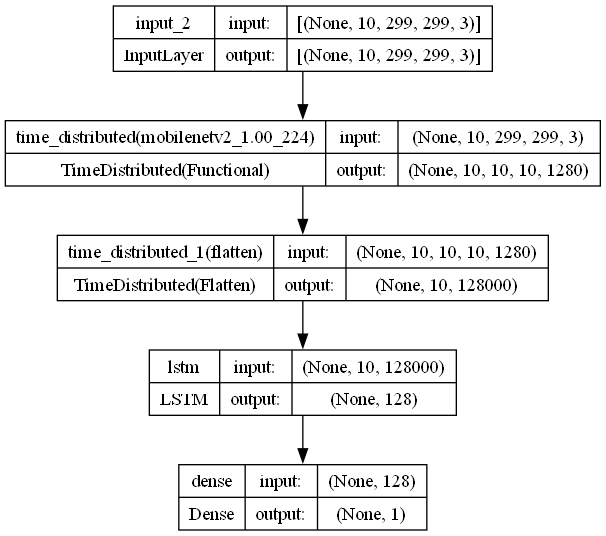}
    \caption{MobileNetV2 + LSTM Architecture}
    \label{fig:model_arch}
\end{figure}

A Long Short-Term Memory (LSTM) layer was integrated with MobileNetV2 to extend the model's capability to handle temporal patterns. While MobileNetV2 effectively extracts spatial features from individual frames, the LSTM layer captures temporal dependencies, enabling the recognition of evolving emotional expressions over time. This hybrid architecture, combining MobileNetV2 and LSTM (Figure \ref{fig:model_arch}), allows for dynamic emotion recognition, capturing subtle variations that static models may overlook.

The MobileNetV2 + LSTM architecture was evaluated on its ability to detect multiple emotional states—engagement, boredom, confusion, and frustration. Among these, engagement consistently outperformed other emotions, achieving the highest accuracy, precision, and recall. Notably, participant feedback corroborated these findings, with engagement predictions perceived as the most reliable and trustworthy, underscoring the alignment between quantitative performance and subjective interpretation.

\subsubsection{Training and Optimization}
The training process for the MobileNetV2 + LSTM architecture was carefully designed to ensure consistent performance across varied conditions. The dataset was split into training, validation, and test sets using stratified sampling to ensure a balanced representation of all emotional states. This approach minimized bias and improved the model's generalization ability.
The model was trained using a binary cross-entropy loss function, treating each emotional state as an independent binary classification task to support multi-label learning. The Adam optimizer was employed for its adaptive learning rate, facilitating faster convergence and training stability. The learning rate was fine-tuned to $1 \times 10^{-4}$ based on validation performance, and early stopping was implemented to prevent overfitting.
Class imbalance was addressed by calculating class weights using the inverse frequency of each emotion, ensuring that less frequent emotional states contributed proportionally during training. Key hyperparameters included:

\begin{itemize}
    \item \textbf{Batch Size:} Set to 16 to balance memory efficiency with stable weight updates.
    \item \textbf{Epochs:} A maximum of ten epochs with early stopping based on validation loss.
    \item \textbf{Class Weights:} Calculated using:
    \[
        w_i = \frac{N}{k \cdot n_i}
    \]
    where $w_i$ is the weight for class $i$, $N$ is the total number of samples, $k$ is the number of classes, and $n_i$ is the number of samples for class $i$.
\end{itemize}

Training checkpoints were saved at each epoch based on validation loss, ensuring the best-performing model was retained. This design optimized the architecture's ability to learn effectively while maintaining stability across emotional classes.

\subsubsection{Evaluation and Results}
The proposed architecture integrates MobileNetV2 for spatial feature extraction and an LSTM layer for temporal sequence modeling. The Time Distributed Wrapper applied to MobileNetV2 ensures frame-wise processing while preserving temporal order, enabling the LSTM layer to detect evolving patterns in video-based emotion analysis. With a sigmoid activation function, the final dense output layer produces probability scores for each emotional state, facilitating multi-label classification. The model's performance was evaluated using key metrics:

\begin{itemize}
    \item \textbf{Accuracy:} Ratio of correct predictions to total instances.
    \item \textbf{Precision:} Ratio of true positives to all positive predictions, minimizing false positives.
    \item \textbf{Recall:} Ratio of true positives to all actual positives, ensuring emotional variations are captured.
    \item \textbf{F1-Score:} Harmonic mean of precision and recall, balancing both metrics.
    \item \textbf{ROC AUC Score:} Ability of the model to distinguish between positive and negative classes across thresholds.
\end{itemize}

Table~\ref{tab:performance_metrics} summarizes the results for each emotional state. Engagement achieved the highest overall performance, with precision of 92\%, recall of 80\%, and an F1-score of 86\%. Frustration and confusion also demonstrated strong performance, while boredom showed slightly lower recall, likely reflecting the subtler nature of its expressions. The ROC AUC scores indicate strong discrimination ability across emotions, with engagement achieving 0.72 and other emotions ranging from 0.67 to 0.80. The model's consistent performance highlights its suitability for real-time applications. Feedback from participants further validated the reliability of engagement predictions, making it the focus of subsequent phases to optimize practical implementations.

\begin{table}[t!]
    \centering
    \renewcommand{\arraystretch}{1.2}
    \caption{Performance Metrics for MobileNetV2 + LSTM by Emotion}
    \label{tab:performance_metrics}
    \resizebox{\textwidth}{!}{
    \begin{tabular}{|l|c|c|c|c|c|}
    \hline
    \textbf{Emotion} & \textbf{Precision (\%)} & \textbf{Recall (\%)} & \textbf{F1-Score (\%)} & \textbf{Accuracy (\%)} & \textbf{ROC AUC} \\ \hline
    Engagement       & 92.0 & 80.0 & 86.0 & 80.0 & 0.72 \\ \hline
    Boredom          & 78.0 & 73.0 & 74.0 & 73.0 & 0.80 \\ \hline
    Confusion        & 86.0 & 87.0 & 87.0 & 87.0 & 0.67 \\ \hline
    Frustration      & 93.0 & 89.0 & 91.0 & 89.0 & 0.74 \\ \hline
    \end{tabular}%
    }
\end{table}

In summary, the MobileNetV2 + LSTM architecture demonstrates a robust approach to video-based emotion detection, effectively integrating spatial and temporal modeling. Its consistent performance across diverse emotional states establishes a strong foundation for real-world applications in adaptive learning and interactive systems.

\subsection{Data Collection}
The data collection phase utilized a structured Human-in-the-Loop (HITL) framework to refine emotion detection models through real-time participant feedback. Three distinct scenarios—Scenario 1 (S1), Scenario 2 (S2), and Scenario 3 (S3)—were designed to examine human interactions with model predictions under varying conditions, focusing on model reliability, error tolerance, and cognitive biases. These scenarios aimed to capture patterns in human-model interaction and support the development of a more reliable engagement detection system.
Each scenario featured unique experimental conditions:

\begin{itemize}
    \item[S1] Conducted initially as a pilot study with 15 participants, S1 evaluated reactions to unaltered model predictions across four emotions: Engagement, Boredom, Confusion, and Frustration. Engagement predictions showed the lowest disagreement rates, prompting the study to focus exclusively on engagement in later scenarios. The remaining participants assessed engagement predictions under similar conditions to establish a baseline measure of perceived model reliability.
    \item[S2] Model predictions were intentionally distorted using the formula \( |100 - x| \), where \( x \) represents the original engagement score. This manipulation encouraged participants to critically assess predictions and provide corrective feedback, enabling analysis of their ability to identify errors and willingness to intervene.
    \item[S3] Participants were informed of the model’s history of poor performance, though predictions remained unaltered. This scenario investigated the impact of perceived reliability on trust, skepticism, and corrective actions, even when the predictions were accurate.
\end{itemize}

\begin{figure}[t]
    \centering
    \begin{subfigure}[t]{0.3\textwidth}
        \centering
        \includegraphics[width=\textwidth]{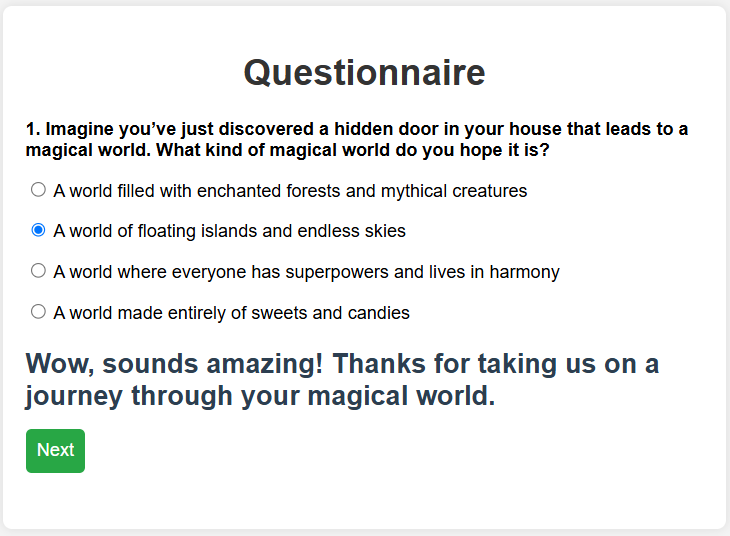}
        \caption{Sample Imaginative Question for S1}
        \label{fig:S1}
    \end{subfigure}
    \hfill
    \begin{subfigure}[t]{0.3\textwidth}
        \centering
        \includegraphics[width=\textwidth]{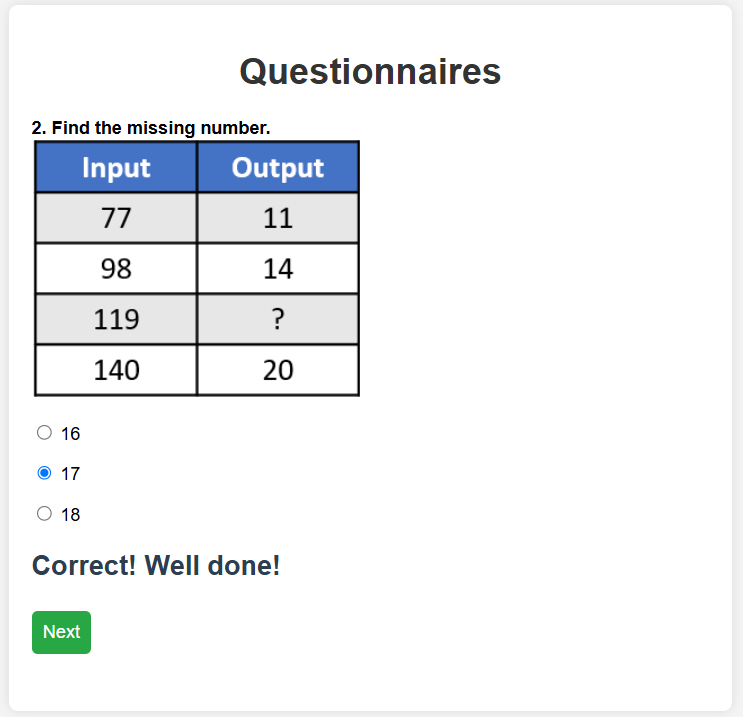}
        \caption{Sample Puzzle Question for S2}
        \label{fig:S2}
    \end{subfigure}
    \hfill
    \begin{subfigure}[t]{0.3\textwidth}
        \centering
        \includegraphics[width=\textwidth]{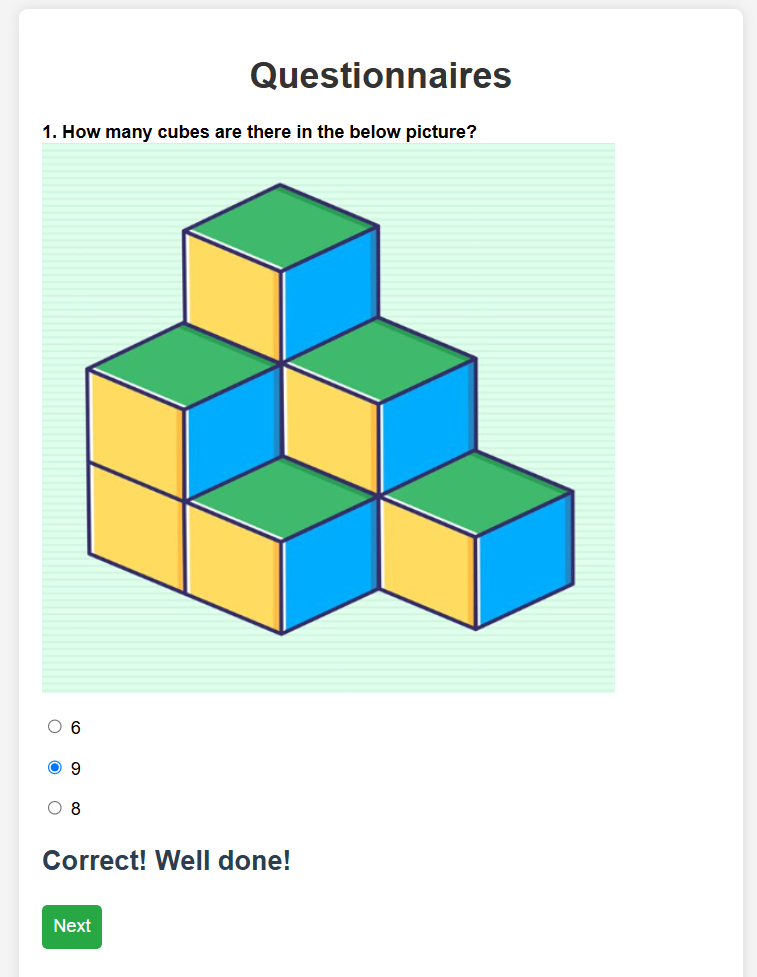}
        \caption{Sample Puzzle Question for S3}
        \label{fig:S3}
    \end{subfigure}
    \caption{Comparison of Questions Across Scenarios}
    \label{fig:scenarios}
\end{figure}

The three scenarios collected qualitative and quantitative data to examine participant agreement, correction behaviors, and the influence of reliability framing on trust and engagement. S1 established a baseline for model reliability, while Scenarios 2 and 3 introduced challenges to elicit critical feedback. Open-ended exploratory questions were used in S1, while S2 and S3 employed structured puzzle-based questions for focused evaluation. Sample displays of these question types are shown in Figures~\ref{fig:S1}, \ref{fig:S2}, and \ref{fig:S3}.

Participant recruitment and instructions were critical components of the data collection process. A total of 33 participants were recruited for S1, with 15 participating in the pilot study before expanding to include the remaining participants. Scenarios 2 and 3 involved 29 participants, as unforeseen circumstances prevented four individuals from continuing. All participants provided informed consent under GDPR guidelines, ensuring transparency regarding the study’s purpose, data usage, and their rights.

Before the study, participants received clear instructions on evaluating the model’s predictions, including agreeing, disagreeing, and providing corrective feedback for inaccuracies. In S3, additional instructions introduced cognitive bias by informing participants of the model’s low reliability to encourage increased skepticism. Scenarios 1 and 2 included neutral instructions, enabling natural interaction with predictions. All participants worked independently in controlled environments to ensure consistency, minimize external influences, and support uniform data collection.

\section{Results and Analysis}
\subsection{Scenario Based Analysis}
The scenario-based analysis examines participant interactions with the model's engagement predictions under varying conditions, providing insights into trust, corrective behaviors, and the influence of cognitive framing.

\subsubsection{Scatter Plots: Alignments and Deviations - } 
\begin{figure}[t] 
    \centering
    \begin{subfigure}[b]{0.32\textwidth} 
        \centering
        \includegraphics[width=\textwidth]{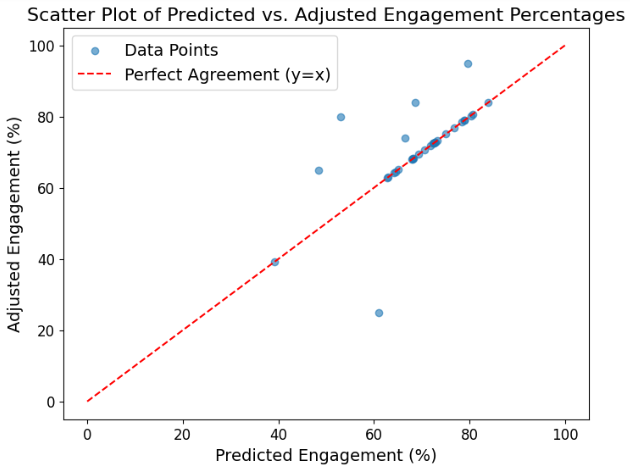} 
        \caption{S1 - Baseline Analysis}
        \label{fig:Scatter_S1}
    \end{subfigure}
    \hfill 
    \begin{subfigure}[b]{0.32\textwidth} 
        \centering
        \includegraphics[width=\textwidth]{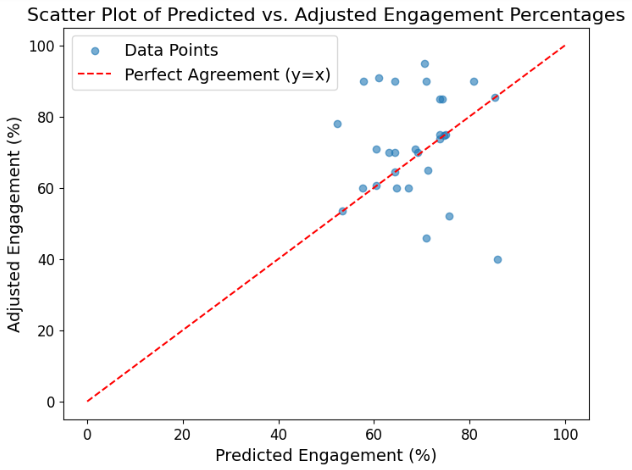} 
        \caption{S2 – Manipulated}
        \label{fig:Scatter_S2}
    \end{subfigure}
    \hfill 
    \begin{subfigure}[b]{0.32\textwidth} 
        \centering
        \includegraphics[width=\textwidth]{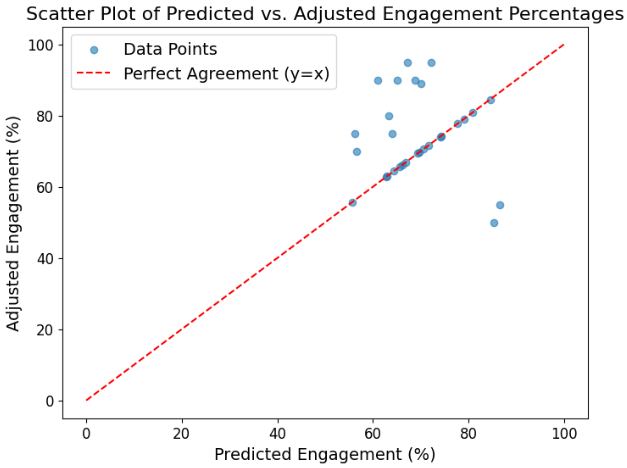} 
        \caption{S3 - Cognitive Bias}
        \label{fig:Scatter_S3}
    \end{subfigure}
    \caption{Scatter plots comparing predicted and adjusted engagement percentages across three scenarios: baseline reliability (S1), manipulated predictions (S2), and cognitive bias (S3).}
    \label{fig:visual_comparison_scatter}
\end{figure}

The scatter plot illustrates the relationship between the predicted and adjusted engagement percentages across the three scenarios. In S1 (Figure~\ref{fig:Scatter_S1}), the points are mostly clustered near the diagonal, showing a strong agreement between the model's predictions and participant adjustments. In S2 (Figure~\ref{fig:Scatter_S2}), we see a significant shift away from the diagonal, as participants actively corrected the flawed predictions, highlighting their engagement and willingness to intervene. The larger adjustments reflect their efforts to correct manipulated predictions. For S3 (Figure~\ref{fig:Scatter_S3}), the scatter points show a moderate deviation, reflecting how the negative framing of the model's reliability led to increased skepticism and adjustments, even though the predictions were accurate. These plots clearly show how trust, accuracy, and cognitive framing influenced participant behavior across the scenarios.

\subsubsection{Density Plots: Distribution Trends - }
\begin{figure}[t] 
    \centering
    \begin{subfigure}[b]{0.32\textwidth} 
        \centering
        \includegraphics[width=\textwidth]{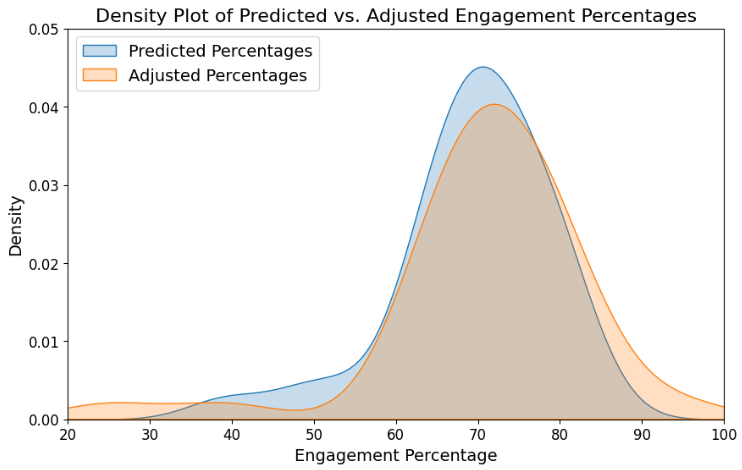} 
        \caption{S1 - Baseline Analysis}
        \label{fig:Density_S1}
    \end{subfigure}
    \hfill 
    \begin{subfigure}[b]{0.32\textwidth} 
        \centering
        \includegraphics[width=\textwidth]{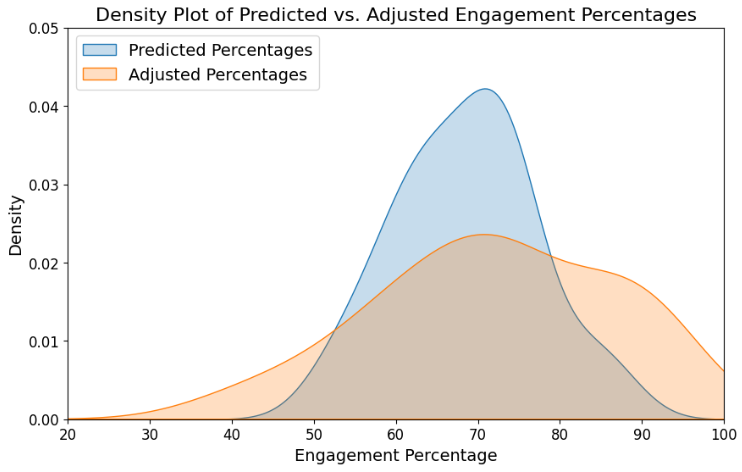}  
        \caption{S2 – Manipulated}
        \label{fig:Density_S2}
    \end{subfigure}
    \hfill
    \begin{subfigure}[b]{0.32\textwidth} 
        \centering
        \includegraphics[width=\textwidth]{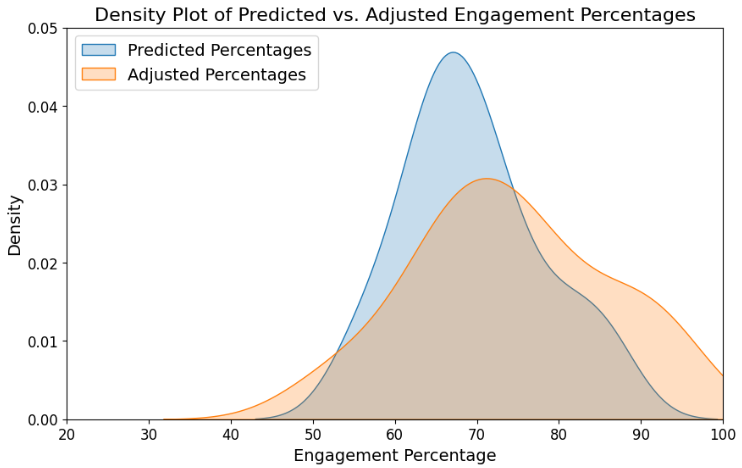} 
        \caption{S3 - Cognitive Bias}
        \label{fig:Density_S3}
    \end{subfigure}
    \caption{Density plots comparing the distributions of predicted and adjusted engagement percentages across three scenarios: baseline reliability (S1), manipulated predictions (S2), and cognitive bias (S3).}
    \label{fig:visual_comparison_density}
\end{figure}

The density plot provides insight into predicted and adjusted engagement percentage distribution. In S1 (Figure~\ref{fig:Density_S1}), the high overlap between predicted and adjusted distributions emphasizes the model's reliability. Participants made minimal adjustments, reinforcing trust in the predictions. In S2 (Figure~\ref{fig:Density_S2}), the density of adjusted percentages shifts significantly, diverging from the manipulated predictions, as participants actively corrected perceived inaccuracies. This divergence underscores their ability to identify and refine flawed outputs. For S3(Figure~\ref{fig:Density_S3}), the density plot reveals partial overlap, with adjusted values skewed higher, suggesting that cognitive bias prompted participants to perceive the model as underestimating engagement. These findings highlight the influence of model reliability and psychological framing on adjustment behaviors.

\subsubsection{Bar Charts: Agreement and Skepticism - } 
\begin{figure}[t] 
    \centering
    \begin{subfigure}[b]{0.32\textwidth} 
        \centering
        \includegraphics[width=\textwidth]{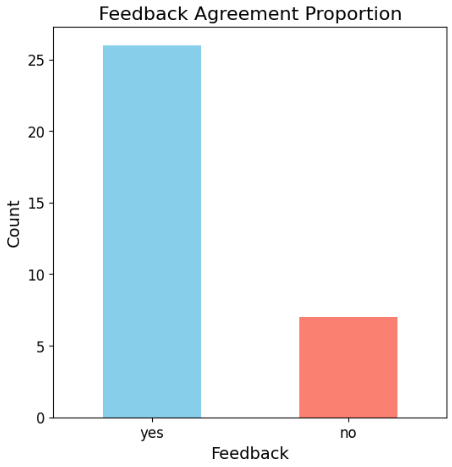} 
        \caption{S1 - Baseline Analysis}
        \label{fig:Bar_S1}
    \end{subfigure}
    \hfill 
    \begin{subfigure}[b]{0.32\textwidth} 
        \centering
        \includegraphics[width=\textwidth]{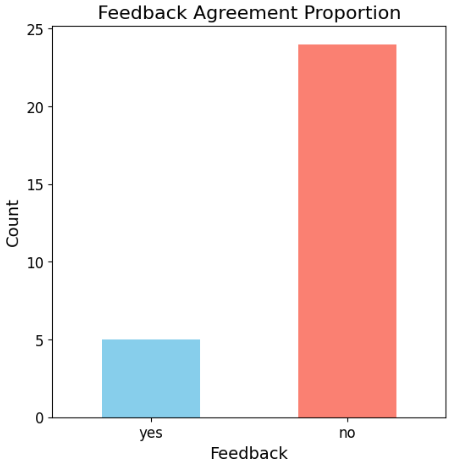}  
        \caption{S2 – Manipulated}
        \label{fig:Bar_S2}
    \end{subfigure}
    \hfill 
    \begin{subfigure}[b]{0.32\textwidth} 
        \centering
        \includegraphics[width=\textwidth]{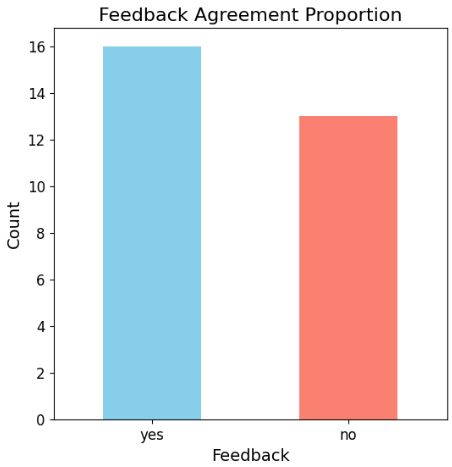} 
        \caption{S3 - Cognitive Bias}
        \label{fig:Bar_S3}
    \end{subfigure}
    \caption{Bar charts illustrating feedback agreement proportions across three scenarios: baseline reliability (S1), manipulated predictions (S2), and cognitive bias (S3).}
    \label{fig:visual_comparison_bars}
\end{figure}

The bar chart illustrates feedback agreement proportions across scenarios, shedding light on participants' trust in the model's predictions. In S1 (Figure~\ref{fig:Bar_S1}), the high proportion of agreement reflects strong trust in unaltered predictions and minimal cognitive strain. In S2 (Figure~\ref{fig:Bar_S2}), the dominant level of disagreement reveals participants' rejection of manipulated predictions and their active correction efforts. S3 (Figure~\ref{fig:Bar_S3}) presents a more balanced distribution of agreement and disagreement, influenced by cognitive framing. While participants remained skeptical due to the negative framing, they still aligned with accurate predictions in some cases, reflecting the interplay between skepticism and trust.

\subsubsection{Box Plots: Variability in Adjustments - }
\begin{figure}[t] 
    \centering
    \begin{subfigure}[b]{0.32\textwidth} 
        \centering
        \includegraphics[width=\textwidth]{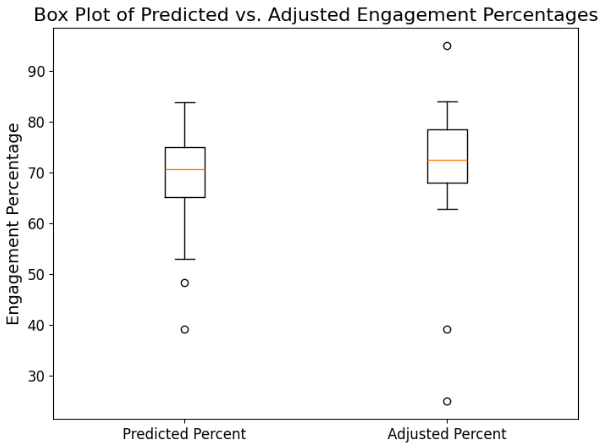} 
        \caption{S1 - Baseline Analysis}
        \label{fig:Box_S1}
    \end{subfigure}
    \hfill 
    \begin{subfigure}[b]{0.32\textwidth} 
        \centering
        \includegraphics[width=\textwidth]{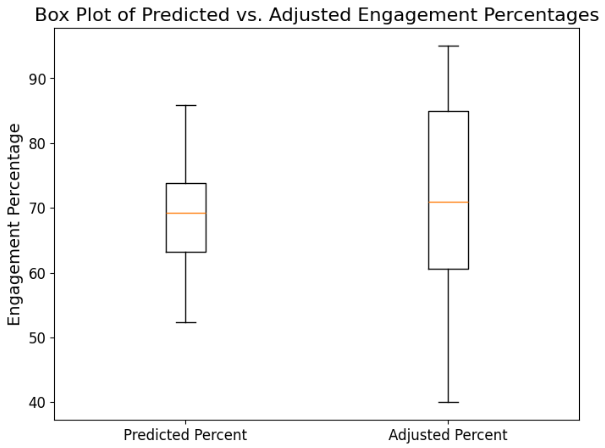}  
        \caption{S2 – Manipulated}
        \label{fig:Box_S2}
    \end{subfigure}
    \hfill 
    \begin{subfigure}[b]{0.32\textwidth} 
        \centering
        \includegraphics[width=\textwidth]{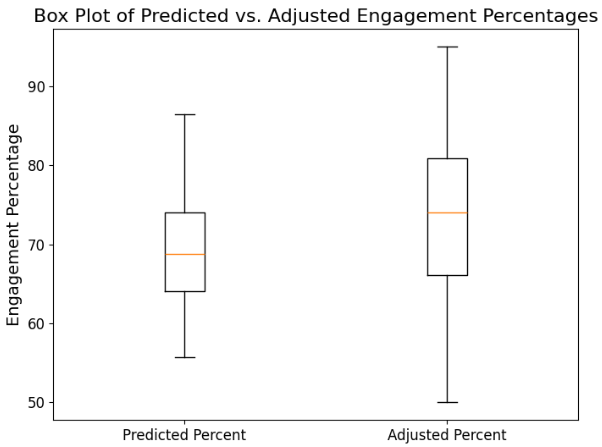} 
        \caption{S3 - Cognitive Bias}
        \label{fig:Box_S3}
    \end{subfigure}
    \caption{Box plots depicting the variability in engagement adjustments across three scenarios: baseline reliability (S1), manipulated predictions (S2), and cognitive bias (S3).}
    \label{fig:visual_comparison}
\end{figure}

The box plot highlights the variability in engagement adjustments across scenarios. In S1 (Figure \ref{fig:Box_S1}), narrow interquartile ranges reflect minimal variability and strong alignment between predictions and adjustments, suggesting high trust and consistency in participant behavior. In S2 (Figure \ref{fig:Box_S2}), wider interquartile ranges indicate substantial corrections made by participants to address the manipulated predictions, showcasing their critical engagement. For S3 (Figure \ref{fig:Box_S3}), the box plot shows moderate variability, reflecting the influence of cognitive bias on adjustment behavior. Participants made more frequent and larger corrections than in S1, as framing effects prompted them to question predictions more actively.

\subsection{Statistical Analysis}
To evaluate the impact of model reliability on participant feedback, statistical tests were conducted to analyze differences across the three scenarios. Key results are summarized in Table~\ref{tab:statistical_analysis}.

\begin{table}[h!]
    \centering
    \setlength{\tabcolsep}{8pt} 
    \renewcommand{\arraystretch}{1.3} 
    \caption{Statistical Test Results and Metrics}
    \label{tab:statistical_analysis}
    \begin{tabular}{@{}llc@{}}
    \toprule
    \textbf{Test} & \textbf{Metric} & \textbf{Value} \\ \midrule
    ANOVA & p-value & 0.0187 \\
    Kruskal-Wallis & p-value & 0.00037 \\
    Chi-square & p-value & 7.41e-06 \\
    Cramér's V & Value & 0.5095 \\ \midrule
    \multirow{3}{*}{Kolmogorov-Smirnov (KS)} 
        & S1 vs. S2 (Predicted/Adjusted) & 0.675 / 0.138 \\
        & S1 vs. S3 (Predicted/Adjusted) & 0.617 / 0.419 \\
        & S2 vs. S3 (Predicted/Adjusted) & 0.951 / 0.791 \\ \midrule
    \multirow{3}{*}{Cohen's d} 
        & S1 vs. S2 & -0.729 \\
        & S1 vs. S3 & -0.546 \\
        & S2 vs. S3 & 0.164 \\ 
    \bottomrule
    \end{tabular}
\end{table}

\begin{itemize}
    \item ANOVA~\cite{st1989analysis} and Kruskal-Wallis Tests~\cite{mckight2010kruskal}: Revealed significant differences in engagement adjustments across scenarios, highlighting the variability introduced by experimental manipulations.
    \item Chi-square Test~\cite{mchugh2013chi} and Cramér’s V~\cite{bergsma2013bias}: Showed a strong association between scenario type and feedback agreement, emphasizing the influence of model reliability on participant trust.
    \item Kolmogorov-Smirnov (KS) Test~\cite{berger2014kolmogorov}: Indicated that S2 (manipulated predictions) differed significantly from Scenarios 1 and 3 in terms of engagement distribution, with a wider spread in adjustments.
    \item Cohen’s d~\cite{peng2014beyond}: Highlighted the largest effect size between Scenarios 1 and 2 ($d = -0.729$), reflecting the behavioral impact of fabricated errors in S2.
\end{itemize}

These tests confirmed that model reliability and framing significantly influenced participant behavior and feedback.

\subsection{Qualitative Feedback Analysis}
The open-ended feedback from participants across the three scenarios provides valuable insights into how model reliability and framing influenced their engagement, perceptions, and emotional responses.

\begin{itemize}
    \item S1: Described as intuitive and straightforward, fostering trust in the model’s outputs and minimizing cognitive strain.
    \item S2: Elicited critical thinking but also frustration due to fabricated inaccuracies, highlighting the cognitive demands of correcting errors.
    \item S3: Balanced skepticism and curiosity, with participants describing the task as engaging and reflective despite the negative framing.
\end{itemize}

Participants favored S3 for its relatable predictions and deeper interaction, while S1 was valued for its reliability and simplicity. S2 was engaging for some but criticized for its inconsistency and the cognitive effort required.

\section{Discussion}
\subsection{Interpretation of Results}
The study reveals the complex interplay between model reliability, participant engagement, and annotation behavior in Human-in-the-Loop (HITL) systems for emotion estimation. The three experimental scenarios provide key insights into human-machine interaction, informing the design of HITL frameworks.

S1 emphasized the role of reliable model predictions in fostering trust and reducing cognitive load. Participants agreed with unaltered predictions, making minimal adjustments and demonstrating alignment between the model’s outputs and human judgment. This reliability enabled annotators to focus on refinement rather than questioning validity, highlighting the importance of consistent predictions for maintaining user confidence.

S2, with intentionally flawed predictions, revealed challenges posed by unreliable models. Participants identified and corrected errors, reflecting critical engagement but at the cost of increased cognitive effort and adjustment variability. For some, this led to frustration, showing that while critical thinking is essential, overly unreliable predictions risk disengagement and reduced trust.

S3 examined cognitive bias by framing the model as historically unreliable despite unaltered predictions. Negative framing led participants to engage more skeptically, balancing doubt with objective evaluation. This reflective engagement highlighted the influence of cognitive framing on participant trust and behavior, even when prediction quality remained unchanged.

The cross-scenario analysis underscores the importance of balancing reliability, cognitive demands, and framing in HITL systems. S1 highlighted the value of reliable predictions, S2 demonstrated the trade-offs of engaging with flawed outputs, and S3 revealed the subtle impact of psychological framing. Together, these findings offer a comprehensive understanding of factors shaping human-machine collaboration.

In conclusion, the study highlights the need for HITL systems that prioritize reliable predictions, manage cognitive demands, and account for framing effects to optimize user trust and engagement. These insights provide a foundation for designing emotion-aware technologies that support diverse user behaviors while ensuring efficiency and collaboration.

\subsection{Post Survey Analysis}
The post-survey examined relabeling behavior, cognitive traits, and the impact of cognitive framing on annotation decisions in HITL systems. By analyzing relabeling patterns, demographics, and self-reported cognitive abilities, it identified factors influencing engagement and trust across scenarios.

Relabeling behavior indicated engagement and critical evaluation. S3, framed as unreliable, had the highest relabeling (13 participants), reflecting the strong influence of framing. Many who relabeled in S3 also did so in S2, which introduced errors, showing a tendency to question outputs under perceived low reliability. S1, with reliable predictions, had the least relabeling, highlighting reliability’s role in reducing intervention. These patterns suggest framing and individual traits like critical thinking shape behavior.

Younger participants (20–25 years) relabeled most actively, likely due to higher engagement and technological familiarity. Their educational and professional backgrounds may foster critical thinking and confidence. Participants aged 30 and above relabeled less, reflecting pragmatism or greater trust. Gender trends showed males as most labelers, emphasizing the need for inclusive designs.

Cognitive traits, particularly critical thinking, influenced relabeling. Among the 13 S3 labelers, 92\% rated their critical thinking abilities as high, showing skepticism and proactive engagement. Frequent AI users also engaged more, suggesting familiarity fosters confidence in evaluating predictions.

Cognitive framing shaped decision-making, with participants relying on intuition and objective cues like video clarity or prediction ambiguity. These findings highlight the interplay of cognitive and emotional factors in human-AI interaction.

Insights from the post-survey underscore the need for HITL systems to accommodate diverse users. Younger participants and critical thinkers engaged more actively, emphasizing transparency and actionable feedback. Age and gender trends highlight the importance of inclusive, adaptable designs.

In conclusion, annotation behavior in HITL systems is shaped by individual traits, external framing, and system design. Relabeling patterns, cognitive tendencies, and demographics underscore the need for HITL frameworks that balance user diversity, cognitive demands, and trust, guiding the design of inclusive and effective emotion-aware systems.

\subsection{Limitations}
This study provides valuable insights into HITL systems for emotion estimation and highlights areas for improvement. While capturing diverse perspectives (33 participants in S1 and 29 in Scenarios 2 and 3), the sample size could be expanded to improve generalizability and uncover subtle patterns in larger populations. The controlled experimental design ensured consistency but may not fully reflect real-world variability. Structured conditions or perceived expectations may have influenced participants’ feedback. Future research in naturalistic settings could offer complementary insights into user interactions with HITL systems in less controlled environments.

S3 demonstrated how explicit information about model reliability affects trust, but real-world applications involve implicit and explicit cues. Longitudinal studies could explore how trust and engagement evolve, providing deeper insights into trust-building mechanisms. Insights from S2’s intentional errors could be extended to study how annotators respond to genuine error distributions in real-world systems, enhancing understanding of probabilistic uncertainties in practice.

Findings in this study are specific to emotion estimation, and results may vary in domains like healthcare or autonomous systems, where task complexity and human-machine interactions differ. Expanding research to other fields could broaden the applicability of these findings. The controlled setting excluded real-world factors like multitasking, time constraints, and variations in annotator expertise, which future studies should consider to enrich the understanding of HITL systems in practical applications.

\section{Future Work}
Future research should prioritize scaling HITL systems, expanding emotional metrics, and refining annotation frameworks. Scaling to larger datasets and participant pools while optimizing workflows for diverse conditions is critical. Techniques like active learning and uncertainty sampling can minimize human intervention by targeting uncertain or error-prone predictions. Expanding emotional metrics to include frustration, boredom, and confusion could enhance system versatility and provide deeper insights into reliability and cognitive framing effects across emotions.

Multimodal integration—incorporating speech, physiological signals, and text could improve annotation accuracy. Developing hybrid models that combine supervised, semi-supervised, and reinforcement learning could refine predictions. Collaborative annotation platforms may improve consistency and leverage shared expertise among annotators. Further exploration of cognitive framing’s impact on trust, as observed in S3, could guide the design of transparent and interactive feedback mechanisms. Longitudinal studies could investigate how annotator behavior evolves with varying model transparency, while real-time visualizations of corrections’ impact on model accuracy could enhance collaboration.

Adapting HITL principles to domain-specific systems like medical diagnostics and autonomous technologies will ensure scalability and effectiveness. Addressing ethical concerns, including data privacy, annotator fatigue, and bias mitigation, is essential to developing fair and inclusive systems.

In summary, advancing HITL systems requires a focus on scalability, diverse emotional metrics, hybrid frameworks, domain-specific challenges, and ethical considerations. These efforts will strengthen human-machine collaboration and drive the development of emotion-aware technologies.

\section{Conclusion}
This study provides key insights into designing Human-in-the-Loop (HITL) frameworks for emotion estimation, focusing on model reliability, cognitive framing, and human behavior. It demonstrates how reliable predictions build trust (S1), unreliable outputs encourage critical engagement but add strain (S2), and cognitive framing impacts trust even with consistent predictions (S3). The results highlight the need for robust models, adaptive workflows, and transparent communication. Reliable predictions ease annotators’ workload, while strategic framing fosters trust and collaboration. Combining human perspectives with quantitative metrics can further refine emotion-aware systems. These findings extend beyond emotion estimation to areas like human-computer interaction, adaptive learning, and medical diagnostics. HITL frameworks can improve collaboration by addressing diverse user needs, promoting inclusivity, and tackling ethical issues such as bias and data privacy. For instance, engagement metrics could personalize learning in education, and in healthcare, HITL systems could enhance AI diagnostics by ensuring trust and precision. HITL frameworks bridge human judgment and machine precision, enabling responsive, context-aware applications. In summary, HITL frameworks offer a path to accurate, adaptive, and human-aligned technologies. This research provides practical guidance for developing systems that empower users and advance emotion-aware technologies, contributing to better human-machine collaboration and enhanced user experiences.

\bibliographystyle{splncs04}
\bibliography{main}
\end{document}